\documentclass[journal=jacsat,manuscript=article]{achemso}

\usepackage[version=3]{mhchem} 
\usepackage[T1]{fontenc}       
\usepackage[separate-uncertainty = true, 
]{siunitx}
\author{M.~Will}
\affiliation{JARA-FIT and 2nd Institute of Physics, RWTH Aachen University, 52074 Aachen, Germany}
\author{M.~Hamer}
\affiliation{School of Physics and Astronomy and Manchester Centre for Mesoscience and Nanotechnology, University of Manchester, Oxford Road, Manchester M13 9PL, United Kingdom}
\author{M.~M\"uller}
\affiliation{JARA-FIT and 2nd Institute of Physics, RWTH Aachen University, 52074 Aachen, Germany}
\author{A.~Noury}
\author{P.~Weber}
\author{A.~Bachtold}
\affiliation{ICFO-Institut de Ciencies Fotoniques, The Barcelona Institute of Science and Technology, 08860 Castelldefels, Barcelona, Spain}
\author{R.~V.~Gorbachev}
\affiliation{School of Physics and Astronomy and Manchester Centre for Mesoscience and Nanotechnology, University of Manchester, Oxford Road, Manchester M13 9PL, United
	Kingdom}
\author{C.~Stampfer}
\email{stampfer@physik.rwth-aachen.de}
\affiliation{JARA-FIT and 2nd Institute of Physics, RWTH Aachen University, 52074 Aachen, Germany}
\alsoaffiliation{Peter Gr\"unberg Institute (PGI-9), Forschungszentrum J\"ulich, 52425 J\"ulich, Germany}
\author{J.~G\"uttinger}
\affiliation{JARA-FIT and 2nd Institute of Physics, RWTH Aachen University, 52074 Aachen, Germany}

\title{High quality factor graphene-based 2D heterostructure mechanical resonator}
\abbreviations{2D,Nb,NbSe$_2$,PMMA, TLS}
\keywords{mechanical resonator, graphene, NbSe2, 2D heterostructures, NEMS, cavity readout}

\begin{document}

%
%

\newpage
\begin{abstract}
	Ultralight mechanical resonators based on low-dimensional materials are well suited as exceptional transducers of minuscule forces or mass changes. However, the low dimensionality also provides a challenge to minimize resistive losses and heating. Here, we report on a novel approach that aims to combine different 2D materials to tackle this challenge. We fabricated a heterostructure mechanical resonator consisting of few layers of niobium diselenide (NbSe$_2$) encapsulated by two graphene sheets. The hybrid membrane shows high quality factors up to 245'000 at low temperatures, comparable to the best few-layer graphene mechanical resonators. In contrast to few-layer graphene resonators, the device shows reduced electrical losses attributed to the lower resistivity of the NbSe$_2$ layer. The peculiar low temperature dependence of the intrinsic quality factor points to dissipation over two-level systems which in turn relax over the electronic system. Our high sensitivity readout is enabled by coupling the membrane to a superconducting cavity which allows for the integration of the hybrid mechanical resonator as a sensitive and low loss transducer in future quantum circuits. 
\end{abstract}
\textbf{Keywords:} mechanical resonator, graphene, NbSe2, 2D heterostructures, NEMS, cavity readout.
\newpage
Van-der-Waals heterostructures based on complementary two-dimensional (2D) materials are a topic of intense research~\cite{Novoselov2016,Liu2016} as they can be used to fabricate tailored electrical and optical devices with superior properties~\cite{Dean2010, Britnell2013, Koppens2014, Cao2015}. While mechanical devices based on individual 2D materials have shown tunable mechanical frequency and high quality factors~\cite{Eichler2011,Lee2013,Weber2014,Song2014,Singh2014,Cast2015,Morell2016,Guettinger2017}, 
the suitability of 2D heterostructures for mechanical resonators has not been explored so far. Another interesting question is how mechanical vibrations will interact with exotic states in encapsulated 2D materials such as 2D superconductors~\cite{Cao2015,Xi2015} or 2D magnets~\cite{Lee2016, Chittari2016}.

In order to use heterostrucutres for mechanical applications, it is crucial that the mechanical quality factor is not significantly degraded by interlayer friction forces between the materials. So far, a detailed understanding of the energy dissipation in 2D mechanical resonators down to low temperatures is missing, despite various calculations~\cite{Seoanez2007,Kim2009,vonOppen2009} and experiments~\cite{Chen2009,Zande2010,Song2012,Takamura2016}. On the one hand calculations on nanometer sized few-layer graphene resonators suggest the importance of interlayer friction forces~\cite{Kim2009}, on the other hand the highest quality ($Q$) factors were reported in multilayer graphene resonators~\cite{Song2014,Guettinger2017}. Experimental limitations in studying the intrinsic mechanical dissipation are imposed by parasitic edge modes in doubly clamped devices~\cite{Zande2010,Takamura2016}, the dependence of spectral measurements on amplitude fluctuations and frequency fluctuations~\cite{Miao2014,Guettinger2017} and the challenge of a non-invasive readout technique down to milikelvin temperatures~\cite{Guettinger2017}.

In this work we demonstrate a high quality van-der-Waals heterostructure mechanical resonator based on few-layer niobium diselenide (NbSe$_2$) encapsulated by graphene. The membrane motion is sensitively probed by capacitive coupling to a superconducting microwave cavity. This readout technique is ideal to study the energy decay as it allows for time resolved ring-down measurements which are independent of frequency fluctuations~\cite{Guettinger2017, Weber2016}. We demonstrate a high mechanical quality factor over 240'000 which compares favorably to the best multilayer graphene resonators~\cite{Eichler2011,Weber2014,Singh2014, Guettinger2017}. Furthermore, by pulling the membrane electrostatically, the mechanical quality factor changes less compared with pure graphene resonators which can be explained by a reduced electrical resistance. By analyzing the low temperature dependence of the intrinsic energy decay we find an anomalous temperature dependence in the highest Q resonators of both types. This finding can be well understood by modeling the dissipation with two-level fluctuators that mediate the dissipation of mechanical energy to the electronic bath~\cite{Golding1978,Phillips1987, Parpia2008}.

\begin{figure*}
	\centering
	\includegraphics{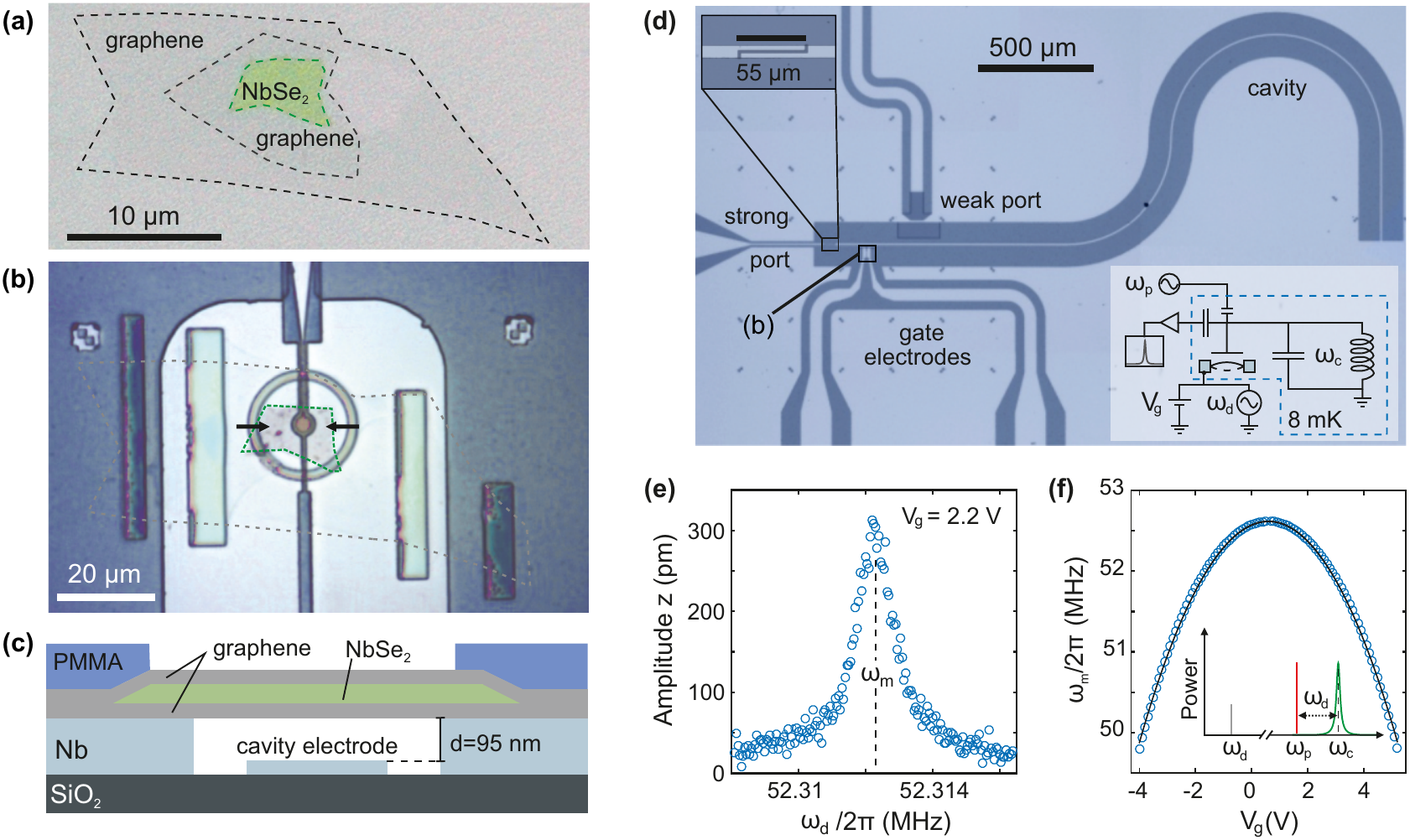}%
	\caption{
		(a) Microscope image of the graphene/NbSe2/graphene assembly on the polymethyl methacrylate (PMMA) membrane before transfer to the microwave circuit. The single-layer nature of the graphene sheets is confirmed by Raman spectroscopy (see SI Section 2). The NbSe$_2$ flake consists of \SIrange{3}{4}{} layers according to optical contrast measurements. (b) Optical microscope image of the hybrid membrane after transfer onto the niobium (Nb) electrodes (light gray). The membrane is fixed to the electrodes by crosslinked PMMA stripes and a ring shaped structure around the suspended device. (c) Schematic cross-section of the mechanical resonator along the black arrows in (b). The gate electrode allows to apply a potential between the membrane and the cavity electrode, displacing the membrane. The initial distance between the cavity electrode and the membrane is \SI{95}{\nano\metre}. (d) Optical contrast image of the entire chip, showing the superconducting cavity including a weak and a strong coupling port and the two electrodes contacting the membrane. Inset: Schematic of the measurement setup. 
		A microwave source with frequency $\omega_\mathrm{p}$ is connected to the weak port and the cavity field is detected over the strong port at the cavity resonance frequency $\omega_\mathrm{c}$. A static membrane-cavity voltage V$_\mathrm{g}$ and a AC mechanical drive voltage at frequency $\omega_\mathrm{d}$ is applied at the membrane. (e) Exemplary spectral measurement of the mechanical amplitude as a function of $\omega_\mathrm{d}$ with the extracted mechanical resonance frequency $\omega_\mathrm{m}$. (f) Resonant frequency $\omega_\mathrm{m}$ of the hybrid membrane as a function of V$_\mathrm{g}$. For the measurement we drive with a constant force $F_\mathrm{d}=\SI{4.46e-14}{\newton}$ (see SI Section 2). The inset shows that we pump the cavity with a red detuned pump tone such that $\omega_\mathrm{p} = \omega_\mathrm{c}-\omega_\mathrm{d}$.}
	\label{f:figure 1} 
\end{figure*}

The optical image of the device in Figure~\ref{f:figure 1} shows the graphene-NbSe$_2$ hybrid membrane, which is capacitively coupled to a superconducting niobium (Nb) cavity. The heterostructure is assembled in an argon atmosphere to avoid contamination and degradation of the NbSe$_2$ layer~\cite{Cao2015}. The membrane consists of few layers of NbSe$_2$ that are encapsulated by single-layer graphene flakes on each side (see Figure~\ref{f:figure 1}a). Alternatively, it is also possible to use hBN crystals or other stable 2D materials for the encapsulation. The main reason to use graphene is the high mechanical strength of the material and the minimal added effective mass by monolayer encapsulation. Another beneficial aspect of graphene is the ease to find single-layer graphene. We expect the NbSe$_2$ layer to induce superconductivity in the graphene due to the proximity effect. However, as the right contact showed an infinite resistance, we were not able to observe superconductivity in transport measurements.
As previously demonstrated with few-layer graphene membranes~\cite{Weber2014}, the encapsulated membrane is transferred on top of a pre-patterned Nb cavity and clamped by cross-linking the polymethyl methacrylate (PMMA) membrane that was used to transfer the 2D membrane (Figure~\ref{f:figure 1}b,c). The membrane is contacted over the left electrode which allows to apply a static gate voltage $V_\mathrm{g}$ and a resonant driving voltage $V_\mathrm{d}$ with respect to the grounded cavity electrode (Figure~\ref{f:figure 1}c,d). 
The as-fabricated device has an initial separation of $d=\SI{95}{\nano\metre}$, which is weakly reduced by \SI{8}{\nano\metre} for the maximum applied voltage of $ V_\mathrm{g}=\SI{-4}{\volt}$ in this work. This static displacement is inferred from the shift of the cavity frequency $\omega_\mathrm{c}/2\pi\approx\SI{7.5}{\giga\hertz}$ as a function of $V_\mathrm{g}$ due to the change of the cavity capacitance~\cite{Weber2016}. 	
Tunable mechanical resonance frequencies, low mass and high mechanical quality factors are key assets of mechanical resonators based on graphene. We show that this quantities can be maintained also in graphene-based heterostructure membranes. Figure~\ref{f:figure 1}e shows the spectral response of the hybrid resonator to a driving tone with a resonance frequency $\omega_\mathrm{m}/2\pi > 50$~MHz. This frequency is comparable to drums with graphene membranes of similar radius to this $r = \SI{1.7}{\micro\metre}$. The mechanical vibration is probed by the capacitive coupling between the electromagnetic pump field in the cavity and the mechanical motion~\cite{Weber2014}.
The pump field is injected into the cavity over a weakly coupled port and the scattered field is detected over a strongly coupled port and amplified by a low noise amplifier at the \SI{3.5}{\kelvin} stage of the cryostat (Figure \ref{f:figure 1}d). For all measurements we use a red-detuned pump tone such that the anti-Stokes scattered field becomes resonant with the cavity frequency (see inset Figure~\ref{f:figure 1}f). The coupling is characterized by the so called single photon-phonon coupling constant $g_0/2\pi\approx \SI{6.1}{\hertz}$, which we estimate in SI~Section 2. We use sufficiently low pump fields such that we can neglect optomechanical backaction.
In order to evaluate the mass of the membrane we measure the $V_\mathrm{g}$ dependence of $\omega_\mathrm{m}$ (see Figure~\ref{f:figure 1}f). The decrease of $\omega_\mathrm{m}/2\pi$ with increasing $|V_\mathrm{g}|$ is a hallmark of an electro-mechanical resonator under tension and can be attributed to a softening of the mechanical potential by the electric field. For the fundamental mode of a circular resonator the gate voltage dependent mechanical resonance frequency can be modeled by~\cite{Weber2014}
\begin{equation}
	\omega_\mathrm{m}(V_\mathrm{g})=\sqrt{\omega_0^2-\frac{0.271\epsilon_0\pi r_\mathrm{g}^2}{d^3 m_\mathrm{eff}}\left(V_\mathrm{g}-V_0\right)^2}.
\end{equation}
Here $\omega_0/2\pi$ is the maximum resonance frequency at the mechanical charge neutrality point ($V_\mathrm{0}=\SI{0.6}{\volt}$), $\epsilon_0$ is the electric constant, $m_\mathrm{eff}=0.27 \pi r^2 \rho_\mathrm{2D}$ the effective mass of the fundamental mode, $\rho_\mathrm{2D}$ the two dimensional mass density of the membrane and $r_\mathrm{g} = \SI{1}{\micro\metre}$ the radius of the gate electrode. From a fit to the data we obtain $m_\mathrm{eff} = \SI{37.7}{\femto\gram}$ and $\omega_0/2\pi =\SI{52.6}{\mega\hertz}$. 
Given the radius of the membrane, this effective mass corresponds to the mass of roughly 20 layers of graphene with $\rho_\mathrm{gr} = \SI{0.76}{\femto\gram\per\micro\metre\squared}$. Considering that one layer of NbSe$_2$ ($\rho_\mathrm{NbSe2} = \SI{4.1}{\femto\gram\per\micro\metre\squared}$) has the mass of 5.2 graphene layers, the mass is in agreement with a heterostructure composed of three layers of NbSe$_2$ encapsulated by two layers of graphene and some extra mass that might be attributed to polymer residues from the fabrication. From $\omega_0$ we can estimate the initial tension of the circular membrane~\cite{Weber2014}
\begin{equation}
	\epsilon \approx \frac{\omega_0^2 m_\mathrm{eff}}{4.92 E_\mathrm{2D}} = \SI{0.08}{\percent},
\end{equation}
with $E_\mathrm{2D}=n_\mathrm{gr}E_\mathrm{2D,gr}+n_\mathrm{NbSe_2}E_\mathrm{2D,NbSe_2}$ the combined two-dimensional Young's modulus. Here $n_\mathrm{gr (NbSe_2)}$ = 2 (3) is the number of graphene (NbSe$_2$) layers and $E_\mathrm{2D,gr (NbSe_2)} = \SI{340}{\newton\per\meter}$ $(\SI{144}{\newton\per\meter})$ is the elastic stiffness of graphene (NbSe$_2$)~\cite{Lee2008,Barmatz1975}.

\begin{figure}
	\centering
	\includegraphics{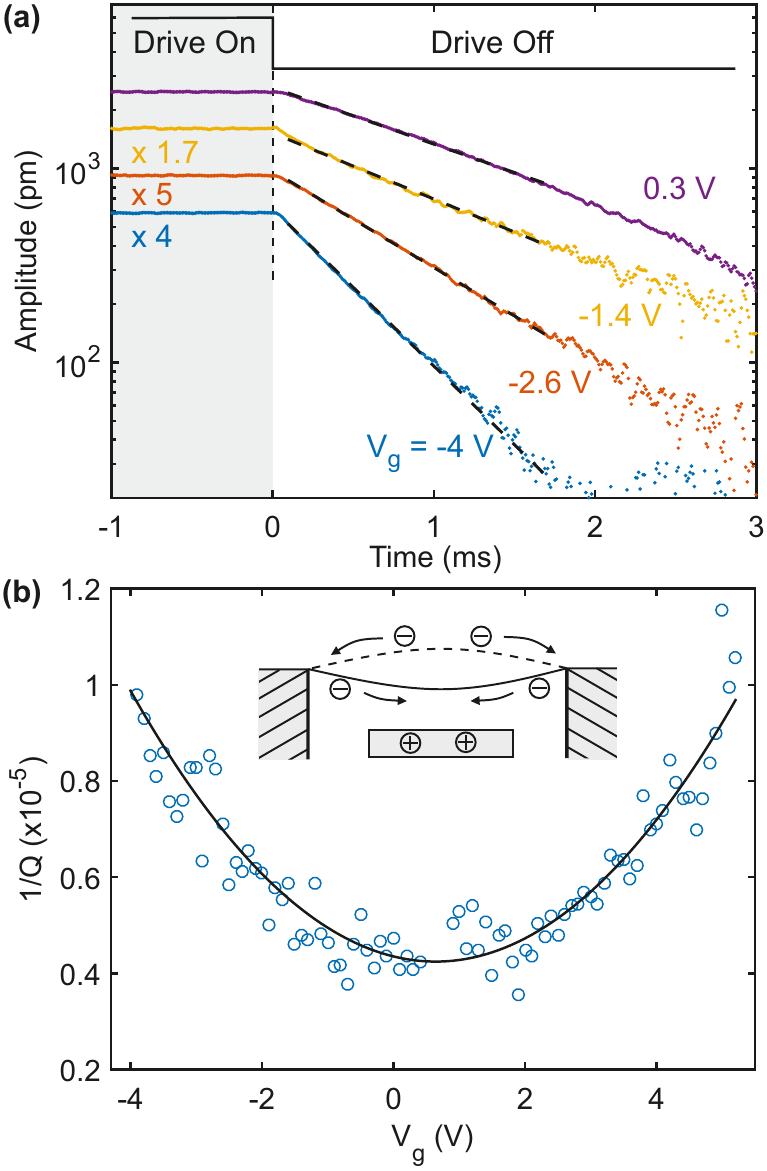}%
	\caption{
		(a) Energy decay measurements at various $V_\mathrm{g}$. The resonator drive is stopped at $t=\SI{0}{\second}$ which leads to a subsequent decay of the mechanical amplitude. The plotted traces are the results of 10000 averaged energy decay measurements. The highest quality factors are obtained at low gate voltage with {values up to $Q=245000$} at $V_\mathrm{g} =\SI{0.3}{\volt}$. (b) $V_\mathrm{g}$ dependence of the inverse quality factor of the membrane (blue). Electronic dissipation is induced over capacitive displacement currents to compensate for the mechanical motion (schematic inset). The fitting parameter for the electronic dissipation is an effective resistance with $R_\mathrm{eff}=\SI{50}{\ohm}$ (black line, see text). Every point is the average of three measurements.}
	\label{f:figure 2}
\end{figure}

In order to probe the mechanical energy dissipation independent of frequency noise and mechanical nonlinearities, we perform ring-down measurements (see Figure~\ref{f:figure 2}). For this measurement, the mechanical resonator is driven to a constant amplitude until the drive is stopped at time $t=\SI{0}{\second}$ and the decay of the amplitude is recorded as a function of time (Figure~\ref{f:figure 2}a). From the exponential energy decay $\propto e^{-\Gamma_\mathrm{decay} t/2}$ the decay rate $\Gamma_\mathrm{decay}/2\pi=\SI{214}{\hertz}$ is extracted at V$_\mathrm{g}=\SI{0.3}{\volt}$. The corresponding quality factor $Q=\omega_\mathrm{m}/\Gamma_\mathrm{decay}={245'000}$ is among the highest measured so far in mechanical resonators based on 2D materials~\cite{Singh2014,Guettinger2017}.

Thanks to reduced electrical loss in our device, a high mechanical Q factor of $\approx 10^5$ is still maintained at increased gate voltage in contrast to few-layer graphene mechanical resonators~\cite{Singh2014, Weber2016}. In Figure~2b the inverse mechanical quality factor obtained from energy decay measurements is plotted as a function of gate voltage $V_\mathrm{g}$. The decrease of the quality factor can be attributed to electrical loss induced by capacitive displacement currents due to the motion of the membrane~\cite{Song2012,Weber2016}. By increasing $V_\mathrm{g}$ more charges have to flow to compensate the change of the capacitance induced by the mechanical motion. The resulting loss of mechanical energy can be described by 
\begin{equation}
	\frac{1}{Q_\mathrm{Joule}}=\frac{R_\mathrm{eff}}{m_\mathrm{eff}\cdot\omega_\mathrm{m}}\left(\frac{\partial C_\mathrm{m}}{\partial z}\right)^2(V_\mathrm{g}-V_0)^2,
\end{equation}
with $R_\mathrm{eff}$ an effective resistance of the membrane and $\partial C_\mathrm{m}/\partial z$ the derivative of the capacitance with respect to the deflection $z$, which is obtained over the measured shift of the cavity resonance frequency as a function of $V_\mathrm{g}$.
The data agrees well with with $1/Q = 1/Q_\mathrm{Joule}+1/Q_0$ with an intrinsic dissipation $Q_0=237'000$ and $R_\mathrm{eff}\approx\SI{50}{\ohm}$ (black line in Figure~\ref{f:figure 2}b). This effective resistance is comparable to the sheet resistance of a three layer NbSe$_2$ flake in the normal conducting state~\cite{Cao2015}.\\  
{Despite the apparent absence of superconductivity in the membrane} the Joule dissipation is reduced by at least a factor of four in the heterostructure device compared with graphene-based mechanical resonators (see Table~\ref{t:table1}). 
\begin{table}
	\centering
	\renewcommand{\arraystretch}{1.25}
	\begin{tabular}[clip, width=1\linewidth]{l|c|c}
		Membrane [layers] &  $R_\mathrm{eff}$ &  $Q_0$  \\\hline
		{30 $\times$ graphene} & \SI{480}{\ohm} & $1.3\times 10^5$\\
		{6 $\times$ graphene} & \SI{220}{\ohm} & $10.6\times 10^5$\\
		{2 $\times$ graphene, 3 $\times$ NbSe$_2$}& \SI{50}{\ohm} & $2.37\times 10^5$
	\end{tabular}    
	\caption{Summary of gate dependent dissipation in different resonators (from Refs.~\cite{Weber2016,Guettinger2017}) in comparison with our heterostructure membrane regarding effective resistance and internal quality factor.}
	\label{t:table1}
\end{table}

\begin{figure}
	\centering
	\includegraphics{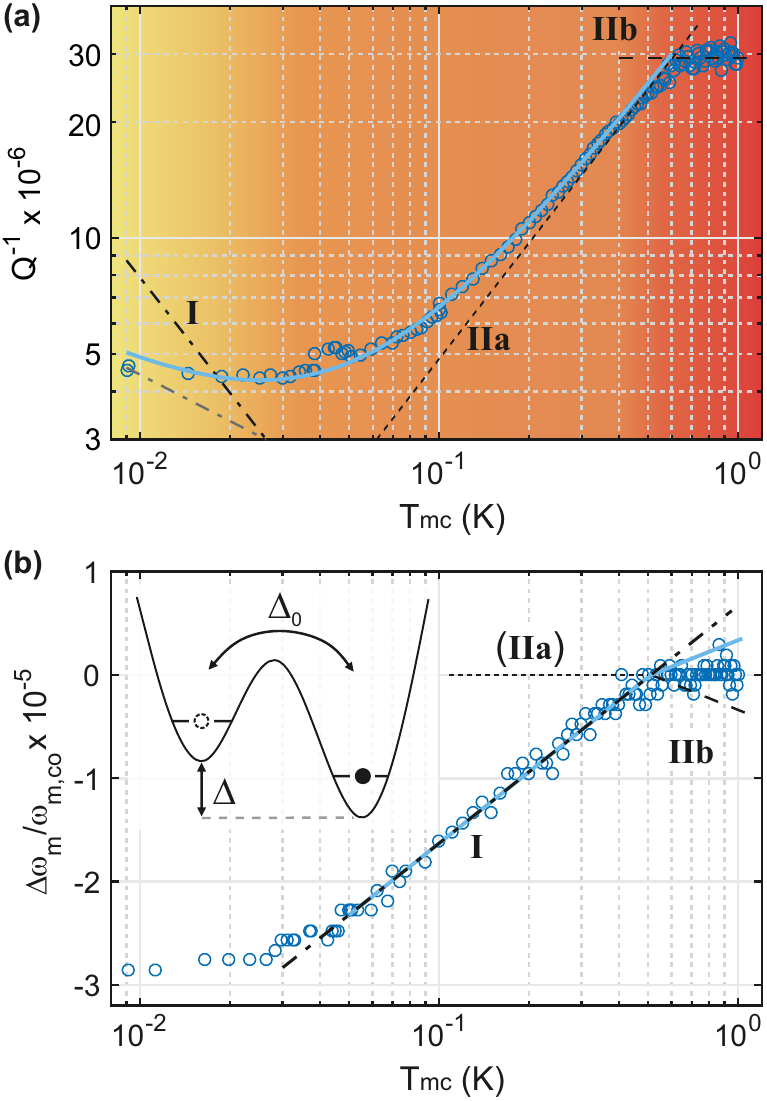}
	\caption{
		(a)	Inverse quality factor $Q^{-1}$ as a function of the cryostat temperature $T_{\text{mc}}$. The color shading highlights three different temperature regimes (yellow, orange, red). The temperature of the mixing chamber (mc) has been measured with a calibrated Magnicon noise thermometer. (b) Relative change of the resonance frequency $\Delta\omega_\mathrm{m}/\omega_\mathrm{m,co}$  as a function of cryostat temperature with respect to the resonance frequency $\omega_\mathrm{m,co}/2\pi = \SI{52.5745}{\mega\hertz}$ at $T_\mathrm{co} = \SI{0.5}{\kelvin}$. The inset depicts a sketch of a TLS with asymmetry $\Delta$ and tunneling coupling $\Delta_0$. In panels (a) and (b) the different black lines correspond to the processes listed in Table~\ref{t:table 2}. The blue line is the sum of the contributions from processes I and II. Note that in panel (a) the grey dot-dashed line with $Q^{-1}=7\times10^{-7}T_{\text{mc}}^{-0.4}$ is used to fit the data in the yellow regime.}
	\label{f:figure 3}
\end{figure}

By analyzing the temperature dependence of $Q$ and $\omega_\mathrm{m}$ in Figure~\ref{f:figure 3} we find evidence that the quality factor is limited by the coupling to two-level systems (TLS). Figure~\ref{f:figure 3}a shows the inverse quality factor as a function of cryostat temperature. The gate voltage was set to $V_\mathrm{g} = \SI{0}{\volt}$ in order to minimize Joule heating losses and probe the intrinsic dissipation. Three different temperature dependencies are identified in the measurements indicated by the yellow, orange and red shading in Figure~\ref{f:figure 3}a. We will focus first on the anomalous crossover at $T_\mathrm{co} \approx \SI{0.5}{\kelvin}$, which cannot be explained by adding two independent dissipation channels with $1/Q_1$ and $1/Q_2$ to $1/Q_\mathrm{tot}=1/Q_1+1/Q_2$. This is because the stronger temperature dependence would dominate at higher temperatures.
It is important to note that this behavior is not related to the heterostructure type of our membrane as a very similar dependence has been observed in a $Q \approx 10^6$ few-layer graphene resonator (see SI Section 3).
Looking at previous low temperature measurements in graphene or nanotube-based resonators~\cite{Jiang2004,Huettel2009,Imboden2014,Takamura2016} a proportionality of $Q^{-1}$ to $T$ or $T^{1/3}$ has been observed individually. However, an increase of the temperature dependence at lower temperature has not been reported before.
We attribute this difference to the higher precision and reduced heating involved in the detection with a superconducting cavity combined with ring-down measurements.
Next we will show that these features can be understood in the framework of two-level fluctuators that are relaxed by the interaction with conduction electrons.

We now recall the model of a two-level fluctuator or tunneling system, which has been proposed to describe low temperature effects in a wide range of isolating and metallic glasses and crystalline resonators~\cite{Anderson1972, Phillips1972, Jaeckle1972, Golding1978, Phillips1987, Kleiman1987, Venkatesan2010, Riviere2011, Esquinazi2013, Lulla2013, Faust2014, Imboden2014}. 
In its simplest form, the tunneling system is defined by an energy asymmetry $\Delta$ and a matrix element $\Delta_0$~\cite{Phillips1987} for tunneling between the two states (see inset in Figure~\ref{f:figure 3}b). For the description of the physics it is often helpful to characterize two-level systems over its energy splitting $E = (\Delta^2+\Delta_0^2)^{1/2}$ and relaxation rate $\tau^{-1}$. Since the resonator interacts with an ensemble of two-level systems, a specific distribution for both $E$ and $\tau^{-1}$ is assumed depending on the exact modeling~\cite{Esquinazi2013}.
The mechanical motion interacts with the TLS over a periodic modulation of the potential. Thereby the TLS can absorb mechanical energy which is subsequently emitted in the phonon or electron bath providing a mechanical energy decay channel. The absorption process of the TLS can be either resonant (I) or over relaxation absorption (II). 

(I) Resonant absorption is dominant at low temperatures where $ k_\mathrm{B}T \approx E \approx \hbar\omega_\mathrm{m}$. The standard TLS model predicts $Q^{-1} \propto \tanh{(\hbar\omega_\mathrm{m}/(2k_\mathrm{B}T))}$ and a resonance frequency shift $\Delta\omega_\mathrm{m}/\omega_\mathrm{m,co} \propto \ln{T}$ for this process~\cite{Phillips1987}. The effects of resonant absorption can be saturated once the mechanical pumping rate gets stronger than the relaxation rate.

(II) Absorption relaxation is dominant for $k_\mathrm{B}T \approx E \gg \hbar\omega_\mathrm{m} $. In this process the mechanical strain fluctuations are affecting the TLS energy mainly over changes in the asymmetry energy $\Delta$~\cite{Phillips1987}. The resulting modulation of $E$ influences the interaction of the TLS with thermal phonons and electrons.
This process gives rise to the peculiar kink in the temperature dependence at $T_\mathrm{co} = \SI{0.5}{\kelvin}$, where the fastest relaxation rate $\tau^{-1}_\mathrm{min}$ equals the mechanical frequency at $T_\mathrm{co}$. In the following we will discuss this process for both temperatures smaller (IIa) and larger (IIb) than $T_\mathrm{co}$. (IIa) For $T < T_\mathrm{co}$ where $\tau^{-1}_\mathrm{min}\ll \omega_\mathrm{m}$ there is a strong temperature dependence of the dissipation with $Q^{-1} \propto T^3$ for phonon relaxation and a linear dependence $Q^{-1} \propto T$ for relaxation over conduction electrons~\cite{Golding1978}. Electronic relaxation leads to a greatly reduced relaxation time which lowers $T_\mathrm{co}$ from a few Kelvin to the sub Kelvin regime for MHz resonators~\cite{Phillips1987}. The associated contribution to the frequency shift is negligible, so that the $T$ dependence of $\omega_\mathrm{m}$ is given by resonant absorption processes involving TLS with $E\approx \hbar \omega_\mathrm{m}$, leading to $\Delta \omega_\mathrm{m}/\omega_\mathrm{m,co}\propto \ln(T)$. (IIb) At $T > T_\mathrm{co}$ where $\tau^{-1}_\mathrm{min}\gg \omega_\mathrm{m}$ the relaxation is faster than the mechanical modulation and the dissipation rate is proportional to $\omega_\mathrm{m}$. The resulting damping is $Q^{-1} \approx \pi C/2$. {The resonance frequency shift due to IIb is negative with $\Delta\omega_\mathrm{m}/\omega_\mathrm{m,co}=-1/2C\ln(T/T_0)$. The combined effect of I and II predicts a reduced but still positive frequency dependence. The positive dependence is not visible in the measurement and might be counteracted by a negative temperature dependence e.g. due to the negative thermal expansion of graphene (see SI Figure~S5). Note that above a few Kelvin the temperature dependence is expected to become negative due to the onset of absorption relaxation by phonons~\cite{Phillips1987}.} 

\begin{table}
	\centering
	\renewcommand{\arraystretch}{1.5}
	\begin{tabular}{c|c|ll}
		Quantity & Process & \multicolumn{2}{c}{Standard TLS model}  \\
		\hline
		& I & $C\ln{(\frac{T}{T_0})}$ &{$\rightarrow C_{\omega_\mathrm{m}}= 1\times 10^{-5}$}\\
		$\frac{\Delta\omega_\mathrm{m}}{\omega_\mathrm{m,co}}$ & IIa & $\approx 0$\\
		& IIb & $-\frac{1}{2}C\ln{(\frac{T}{T_\mathrm{co}})}$  &$\rightarrow T_\mathrm{co} \approx 0.5~$K\\
		\hline
		& I & $\pi C\tanh{[\frac{\hbar\omega_\mathrm{m}}{2k_\mathrm{B}T}]}$ &(saturable)\\
		$Q^{-1}$  & IIa & $\frac{\pi^3C}{24}K^2 \frac{k_\mathrm{B}T}{\hbar\omega_\mathrm{m}}$  &{$\rightarrow K = 0.071$}\\
		& IIb & $\frac{\pi}{2}C$&{$\rightarrow C_Q= 1.9\times 10^{-5}$}\\
	\end{tabular}  
	\caption{Predictions for $\Delta\omega_\mathrm{m}/\omega_\mathrm{m,co}$ and $Q^{-1}$ based on the standard TLS theory in metallic systems~\cite{Phillips1987}. For both quantities the contribution from resonant (I) and relaxation absorption (II) add up. The crossover temperature $T_\mathrm{co}$ separates the different contribution of relaxation absorption for $T\ll T_\mathrm{co}$ (IIa) and for $T\gg T_\mathrm{co}$ (IIb).  $T_0$ depends on the frequency offset where $\Delta\omega_\mathrm{m} = 0$ and is related to the crossing temperature $T_\mathrm{co}$. The parameter $C$ is related to the TLS density and the coupling strength of deformations to the TLS energy splitting. The parameter $K$ is proportional to the conduction electron density and their coupling to TLS.}
	\label{t:table 2} 
\end{table}

A quantitative comparison between theory and experiment is generally difficult and will serve here only to get a qualitative understanding of the parameters. We compare our measurements with the standard TLS theory~\cite{Phillips1987,Esquinazi2013}. We assume that electronic relaxation dominates over phonon relaxation in the measured temperature regime due to the low $T_\mathrm{co}$ (fast relaxation time) and $Q^{-1}\propto T$ below $T_\mathrm{co}$. From the frequency dependence we extract $T_\mathrm{co} \approx 0.5~$K from the kink position and {$C_{\omega_\mathrm{m}} = 1\times 10^{-5}$} from the slope below $T_\mathrm{co}$ (see formulas in Table~\ref{t:table 2}). Note that the frequency offset has been adjusted such that the temperature $T_0$ corresponding to $\Delta\omega_\mathrm{m} = 0$ equals $T_\mathrm{co}$.
{From the saturation of $Q^{-1}$ for $T>T_\mathrm{co}$ we extract $C_Q = 1.9\times10^{-5}$. The difference between $C_{\omega_\mathrm{m}}$ and $C_Q$ might arise from changes in the density of the contributing TLS at the different temperatures.~\cite{Riviere2011}.}
We extract a TLS electron coupling constant of {$K = 0.071$} (dotted line) from the fit of the linear temperature dependence of contribution IIa. The temperature dependence in a high Q graphene device yields comparable values with {$C_{\omega_\mathrm{m}} = 1.6\times 10^{-5}$ and $K = 0.089$} (see SI Section 3). These values are also in the same range that has been measured previously in metallic glasses and polycrystalline metals with $C = 10^{-5}-10^{-4}$ and $K = 0.02-0.2$~\cite{Esquinazi2013,Copper1993}. 
At the lowest temperatures the dissipation has a negative temperature dependence as predicted by theory (see also SI Figure~S3).
{However, the measurement shows a reduced temperature dependence compared to the model (I), which might be explained by insufficient thermalization of the membrane to the mixing chamber plate at the lowest temperatures ($T_\text{mc}<40\,$mK, see Figure~\ref{f:figure 3}b).	
	Please note that the fit merely emphasizes the change of sign in temperature dependence and is not meant to be universally valid.} In the low temperature $Q^{-1}$ dependence of the graphene device we don't observe any signature from process I. This might be explained with a complete saturation of the resonant TLS at the lowest drive amplitudes studied. Furthermore, in contrast to Ref.~\cite{Singh2016} we do not observe a saturation of the damping with increasing drive power but rather an increase in damping attributed to additional decay channels~\cite{Guettinger2017}. 
A deviation from the theoretical prediction of the standard model is not unexpected as it has been observed before in metallic glasses~\cite{Golding1978}. A nonsaturable attenuation might arise from more asymmetric TLS or a modified scattering cross-section~\cite{Copper1993}. Additionally, the effect of individual TLS with $E$ and $\tau^{-1}$ deviating from the assumed contribution is enhanced due to the tiny volume of the membrane~\cite{Remus2009}.

The microscopic mechanism of TLS damping in 2D membranes is unclear so far. A previous calculation on dissipation in graphene and carbon nanotube resonators investigated electrically active TLS in the surrounding of the vibrating structure and concluded that the resulting dissipation will be negligible~\cite{Seoanez2007}. However, residues from fabrication or imperfections in the crystal might still contain mechanically activated TLS~\cite{Imboden2014}. 

TLS damping is not a universal limitation in 2D material based mechanical resonators. In resonators with lower $Q$ factors we observe usually a saturation or reduced temperature dependence at low temperatures. This might be related to clamping induced loss~\cite{Seoanez2007} or loss over parasitic modes~\cite{Zande2010,Takamura2016} which masks the effect of the TLS.

{We have not found any indication for superconductivity in the measured heterostructure device. First, the electric resistance extracted from the dissipation as a function of gate voltage is compatible with normal conducting NbSe$_2$. Second, we would expect the superconductivity to suppress the coupling of the TLS to electrons below $T_\mathrm{c}/2$~\cite{Phillips1987, Lulla2013}, which would lead to a higher kink temperature and increased temperature dependence compared to the measured data.}\\

In conclusion, we presented a mechanical resonator based on van-der-Waals heterostructures of two-dimensional materials. The high mechanical quality, frequency tunability and low electrical resistivity measured in a graphene/NbSe$_2$/graphene resonator exemplifies the opportunities of 2D-heterostructures for mechanical resonators. The enhanced and partially linear temperature dependence of the dissipation is understood with mechanically mediated relaxation of two-level systems by conduction electrons. It might be possible to further improve the mechanical quality by using electrically isolating boron nitride for the encapsulation or reducing polymer residues in the fabrication.
In future experiments, the combination of superconductivity and low mass in a 2D membrane opens the potential for ultra low loss and high sensitivity mechanical resonators. This is because a superconducting mechanical resonator suppresses electrical losses induced by mechanical motion and allows for higher pumping fields in a cavity based readout. Both are important for the exploration of quantum motion and improvements of force sensitivity in atomically thin mechanical resonators~\cite{Weber2014,Song2014,Singh2014,Weber2016}.

\begin{acknowledgement}

We thank Uwe Wichmann for help on the electronics and for fruitful discussions. Support by the Helmholtz Nano Facility (HNF)~\cite{Albrecht2017} at the Forschungszentrum J\"ulich, the Excellence Initiative (RWTH Start-Up grant), the ERC-carbonNEMS and the ERC-GQEMS (GA-Nr. 280140) are gratefully acknowledged. The work is further supported by the ERC advanced grant 692876, the Foundation Cellex, Severo Ochoa (SEV-2015-0522), the grant FIS2015-69831-P of MINECO, and the Fondo Europeo de Desarrollo Regional (FEDER).

\end{acknowledgement}

\begin{suppinfo}

\begin{itemize}
  \item Supplement: Membrane characterization by Raman, calculation of resonator amplitude, data on temperature dependent dissipation in additional devices, effect of thermal expansion on resonance frequency and dependence of dissipation on mechanical drive voltage.
\end{itemize}

\end{suppinfo}

\makeatletter
\providecommand{\doi}
{\begingroup\let\do\@makeother\dospecials
	\catcode`\{=1 \catcode`\}=2 \doi@aux}
\providecommand{\doi@aux}[1]{\endgroup\texttt{#1}}
\makeatother
\providecommand*\mcitethebibliography{\thebibliography}
\csname @ifundefined\endcsname{endmcitethebibliography}
{\let\endmcitethebibliography\endthebibliography}{}

{
\newpage
\section{Supporting information}
\subsection{Characterization of the membrane by Raman}
\begin{figure}
	\centering
	\includegraphics{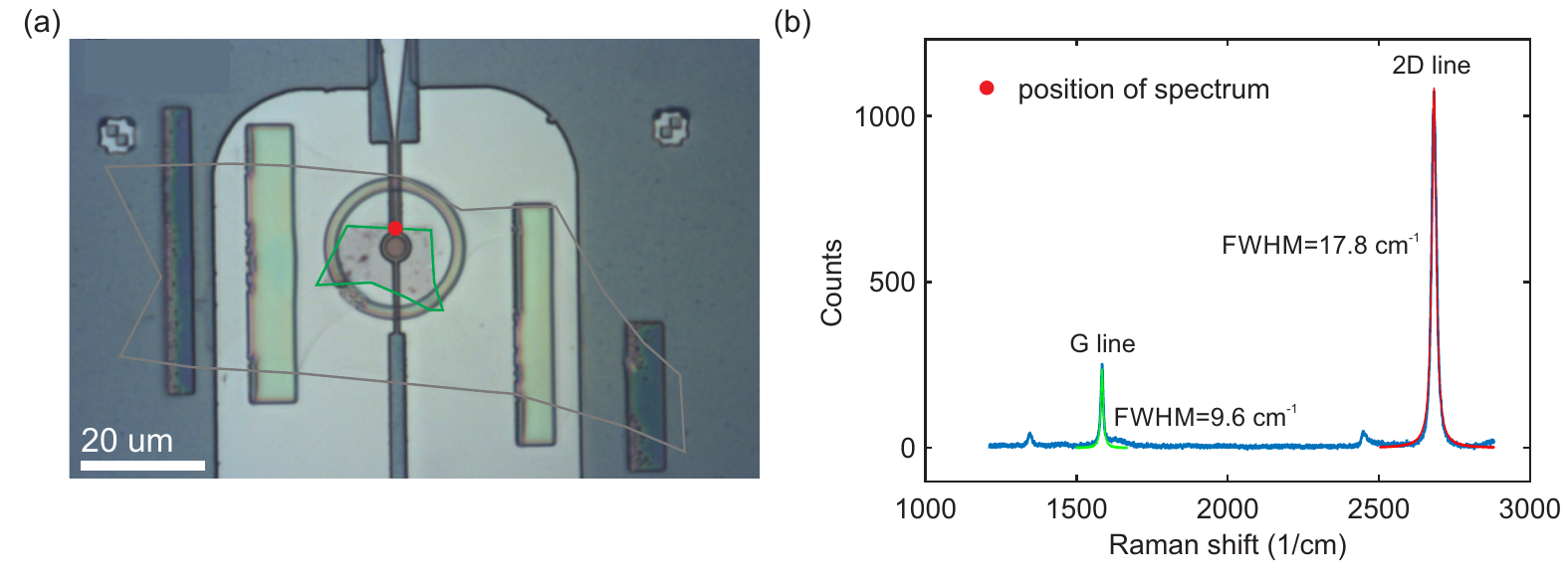}%
	\caption{ (a) Image of the device with the red dot marking the position of the Raman measurement, where both the top and bottom graphene flake are present. (b) Raman measurement of the membrane with single Lorentzian fits of the 2D and G line showing that top and bottom graphene flake are single layer.  
	}
	\label{f:SIFigure1}
\end{figure}
Raman measurements are conducted to determine the number of layers for the encapsulating graphene after the measurement in the dilution refrigerator. Figure~\ref{f:SIFigure1}a shows the position of the suspended region that is investigated with Raman. In Figure~\ref{f:SIFigure1}b the single Lorentzian fit of the 2D peak to the Raman spectrum with a $\mathrm{FWHM}=\SI{17.5}{\per\centi\metre}$ at $\SI{2682.5}{\per\centi\metre}$ indicates single layer graphene. A Raman map of the entire membrane shows that the drum collapsed after the measurement in the dilution refrigerator with only the region around the position marked by the red dot still suspended.
The Raman spectrum of the NbSe$_2$ layer in the drum area is also investigated. However, the resolution of the Raman signal is insufficient to determine the width of the shear mode, which perishes in the Rayleigh peak. The soft mode, A$_\mathrm{1g}$ and E$_\mathrm{2g}$ can be determined and show a Raman shift in the expected region around \SI{240}{\per \centi\metre} (not shown)~\cite{Xi2015_SI}.
\newpage
\subsection{Calculation of the membrane amplitude}
The amplitude $z$ of the membrane can be estimated from the optomechanical coupling as shown in~\cite{Weber2016_SI,Weber2014_SI}:
\begin{equation}
P_\mathrm{out}(\omega_\mathrm{c})=P_\mathrm{in}(\omega_\mathrm{p})\cdot \mathrm{loss}(\omega_\mathrm{p}) \cdot \mathrm{gain}(\omega_\mathrm{c})\cdot \left(\frac{1}{\kappa}\frac{\partial \omega_\mathrm{c}}{\partial z}\right)^2\cdot 2 \langle z^2\rangle\frac{4 \cdot \kappa^2_\mathrm{ext}}{\kappa^2+4(\omega_\mathrm{p}-\omega_\mathrm{c})^2}.
\end{equation}
Here $\kappa$ and $\kappa_\mathrm{ext}$ are the coupling constants of the cavity, $\mathrm{loss}\cdot \mathrm{gain}$ are calculated from the amplifiers and attenuators, $P_\mathrm{in}$ is the power of the pump wave and $\partial \omega_\mathrm{c}/\partial z $ is the change of the cavity resonance frequency with displacement of the membrane. The output power $P_\mathrm{out}$ is measured, whereas the other quantities are extracted from other calibration measurements as explained in the following.

The $\mathrm{loss\cdot gain}$ of the measurement setup inside the cryostat system is \SI{-34}{\decibel}. Adding this to the whole measurement setup outside of the cryostat with a total $\mathrm{loss\cdot gain}=\SI{43.4}{\decibel}$ equals to a total $\mathrm{loss\cdot gain}=\SI{9.3}{\decibel}$. However, for the measurement with the weak port no bandpass filters  with a combined attenuation of \SI{22}{\decibel} are needed resulting in a total $\mathrm{loss\cdot gain}=\SI{31.3}{\decibel}$. This attenuation also leads to a reduction of the drive amplitude. The drive RMS voltage at the sample is adjusted to the applied back gate voltage to avoid high forces resulting in a movement of the membrane. We modify the applied drive voltage $V_\mathrm{d}=10^{(-56.24/10)}\cdot V_\mathrm{d,ZI}/(\SI{100e-6}{}\cdot V_\mathrm{g})$ at the sample so that the drive force is independent of the applied back gate voltage $\hat{F}_\mathrm{d}=\partial_\mathrm{z} C_\mathrm{m}V_\mathrm{g}\sqrt{2}V_\mathrm{d}$.

The read out technique utilizes red sideband pumping of a microwave cavity. The cavity linewidth is determined by the capacitive coupling of the weak and strong port for feed in and read out of the pump frequencies~\cite{Weber2014_SI}. The total cavity linewidth of $\kappa/2\pi\approx\SI{5.9}{\mega\hertz}$ is composed of the capacitive coupling to the two ports $\kappa_\mathrm{strong}/2\pi\approx\SI{2.2}{\mega\hertz}$, $\kappa_\mathrm{weak}/2\pi\approx\SI{1.0}{\mega\hertz}$ and additional internal losses $\kappa_\mathrm{int}/2\pi\approx\SI{2.7}{\mega\hertz}$. The coupling constants are extracted from fits to transmission measurements. All measurements are conducted in a dilution refrigerator with a base temperature of \SI{8.5}{\milli\kelvin} well below the critical temperatures of the Nb cavity ($T_\mathrm{C}  = \SI{9.2}{\kelvin}$).

The single phonon-photon coupling rate $g_0$ is estimated by~\cite{Weber2014_SI}
\begin{equation}
g_0=\frac{\partial \omega_c}{\partial z}\cdot z_\mathrm{zp}=\frac{\omega_\mathrm{c}}{2C_\mathrm{tot}}\frac{\partial C_\mathrm{m}}{\partial z}z_\mathrm{zp} \approx 2\pi \cdot 6.1~\mathrm{Hz}.
\end{equation}
Here $\omega_\mathrm{c}$ is the cavity resonance frequency, $C_\mathrm{tot}=\SI{135}{\femto\farad}$ is the total capacitance of the cavity, z$_\mathrm{zp}=\sqrt{\hbar/(2m_\mathrm{eff}\omega_\mathrm{m})}=\SI{6.5e-14}{\metre}$ is the zero point motion of the resonator and $\partial C_\mathrm{m}/\partial z$ is the derivative of the capacitance between the mechanical resonator and the gate with respect to the center position of the membrane $z$. For $\partial C_\mathrm{m}/\partial z$ we obtain from a plate capacitor estimation $\partial C_\mathrm{m}/\partial z\approx\epsilon_0 \pi R_\mathrm{g}^2/d^2=\SI{3.1e-9}{\farad\per\metre}$, yielding $g_0/2\pi=\SI{5.6}{\hertz}$ in close agreement with $\partial C_\mathrm{m}/\partial z = \SI{3.6e-9}{\farad\per\metre}$ ($g_0/2\pi=\SI{6.5}{\hertz}$) extracted from the change of the cavity resonance frequency by pulling on the membrane with $V_\mathrm{g}$.

Due to the limited coupling we are not able to confirm the estimation of $g_0$ with thermal calibration measurements. However, the coupling is compatible with \SIrange{2}{10}{\hertz} from previous samples with similar geometry~\cite{Weber2016_SI,Weber2014_SI}.
\newpage
\subsection{Temperature dependence of dissipation in a high Q few-layer graphene and an hBN encapsulated graphene device}
\begin{figure}
	\centering
	\includegraphics{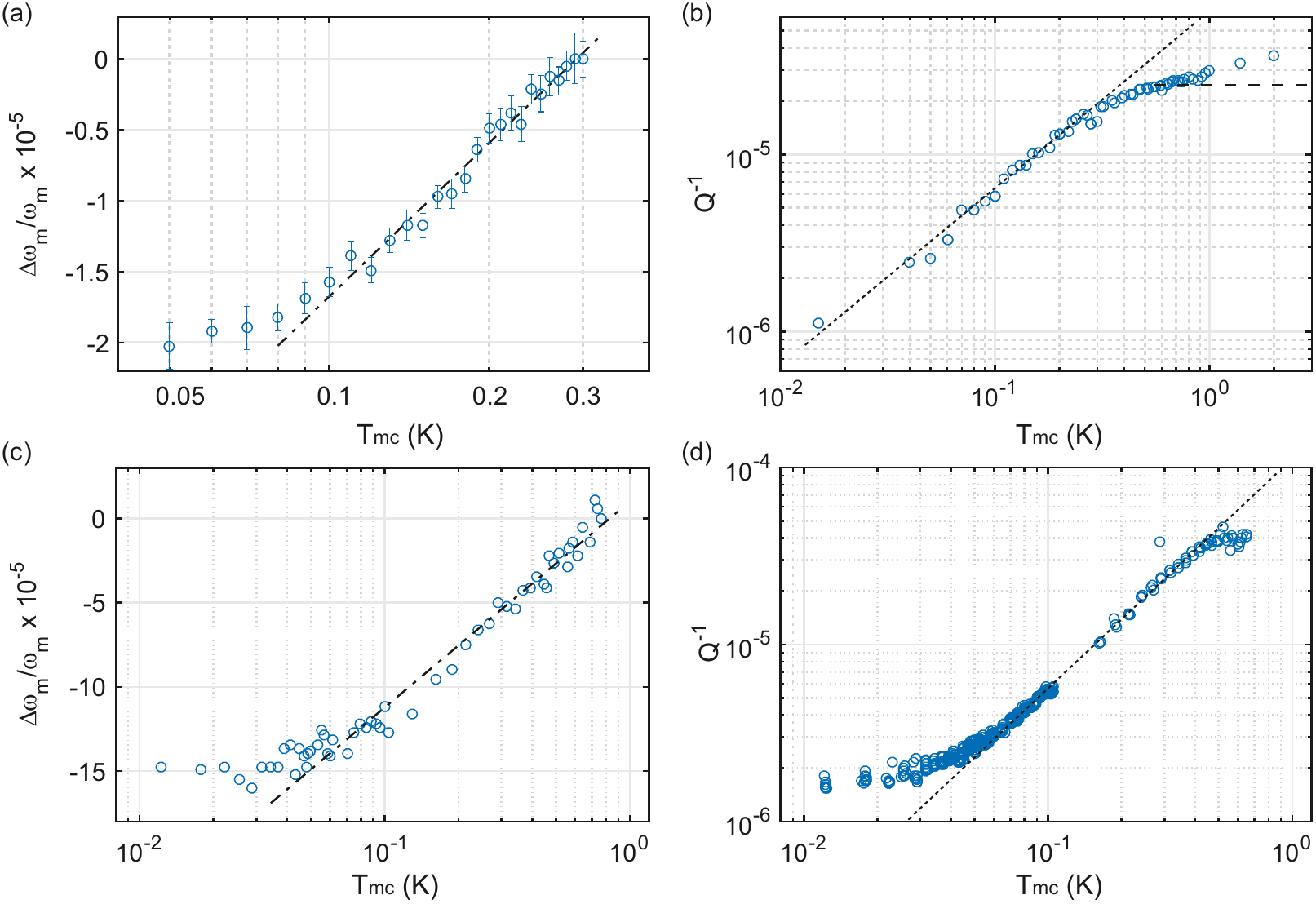}%
	\caption{\label{f:SIFigure2}
		Temperature dependence of $\Delta \omega_\mathrm{m}/\omega_\mathrm{m}$ and $Q^{-1}$  in a high Q few-layer graphene device (a, b) and an encapsulated graphene device (c, d). (a, c) The dotted dashed line corresponds to $\Delta \omega_\mathrm{m}/\omega_\mathrm{m} = C\ln(T/T_\mathrm{0})$ with $C_{\omega_\mathrm{m}} = 1.56\times 10^{-5}$ extracted from a linear fit between \SI{100}{\milli\kelvin} and \SI{300}{\milli\kelvin} (a) and $C_{\omega_\mathrm{m}} = 5.31\times10^{-5}$ from a fit between \SI{70}{\milli\kelvin} and \SI{750}{\milli\kelvin} (c). (b, d) Temperature dependence of $Q^{-1}$ measured with the ring-down method. The dashed line in (b) corresponds to the saturation expected in the standard tunneling model $C_{\omega_\mathrm{m}} = 1.56\times 10^{-5}$ and the dotted line is a fit between \SI{15}{\milli\kelvin} and \SI{240}{\milli\kelvin} to $Q^{-1} = \frac{\pi^3C}{24}K^2 \frac{k_\mathrm{B} T}{\hbar\omega_\mathrm{m}}$. (d) The dotted line represents a fit $\propto T^{1.3}$ between \SI{40}{\milli\kelvin} and \SI{450}{\milli\kelvin}.
	}
\end{figure}
Figure~\ref{f:SIFigure2} shows the temperature dependence of $\Delta \omega_\mathrm{m}/\omega_\mathrm{m}$ and $Q^{-1}$ in a high Q few-layer graphene resonator previously studied in Refs.~\cite{Weber2016_SI,Guettinger2017_SI} and an hBN/graphene/hBN resonator. The resonance frequency change $\Delta\omega_\mathrm{m}/\omega_\mathrm{m}$ in Figure~\ref{f:SIFigure2}a is extracted from thermal motion measurements because the frequency shift is not resolved in the ring-down measurements. Above \SI{100}{\milli\kelvin} the data follows a logarithmic temperature dependence expected from resonant interaction with TLS. The deviation below \SI{100}{\milli\kelvin} is at least partially attributed to heating of the membrane due to the increased cavity pump power needed to resolve the thermal motion~\cite{Weber2016_SI}. The shift in resonance frequency shown in Figure~\ref{f:SIFigure2}c also shows a deviation from the otherwise logarithmic behaviour below \SI{50}{\milli\kelvin} which is connected to an insufficient thermalization of the device to the mixing chamber plate as mentioned in the main text. The extracted value for $C_{\omega_\mathrm{m}}= 5.31\times 10^{-5}$ is in the expected range.\\
The $Q^{-1}$ temperature dependence in Figure~\ref{f:SIFigure2}b shows again a crossover from a stronger to a weaker temperature dependence upon increasing the temperature. The dashed line corresponds to the saturation level $Q^{-1} = \pi/2 \times C_{\omega_\mathrm{m}}$ expected from the standard TLS model with $C_{\omega_\mathrm{m}}= 1.56\times 10^{-5}$ from Figure~\ref{f:SIFigure2}a. However, as with the heterostructure device this plateau does not describe the data well. Besides having a too small value the measurement does not show an immediate saturation but rather a dependence of $Q^{-1} \propto T^{1/3}$ above $T_\mathrm{co} \approx \SI{300}{\milli\kelvin}$. Such a dependence has been observed in several graphene based resonators above \SI{1}{\kelvin}~\cite{Imboden2014_SI}. To still find an estimate for $K$ it was calculated using the previously extracted value for $C_{\omega_\mathrm{m}}$ yielding $K = 0.089$. Another difference compared to the heterostructure device in the main text is the absence of a clear sign for a resonant dissipation contribution to $Q^{-1}$ at the lowest temperatures.
In the hBN/graphene/hBN device the effect of saturation due to resonant absorption and insufficient thermalization in combination with a lower resonance frequency of around \SI{35}{\mega\hertz}, which should shift the position of the change in sign to lower temperatures, is not distinguishable.\\
It will be interesting to investigate this regime in more devices and for higher mechanical modes to gain more insight.\\
The inverse Quality factor dependence of the encapsulated device (Figure~\ref{f:SIFigure2}d) can be described by a fit $\propto T^{1.3}$ for $T_\text{mc}>\SI{50}{mK}$ which still indicates, that the dissipation is connected to electrons rather than phonons.
\newpage
\subsection{Dependence of the dissipation on mechanical drive voltage}
In the measured range the heterostructure device does not show the expected saturation of the dissipation when increasing the drive voltage at $T_\mathrm{mc}=\SI{8}{\milli\kelvin}$ but rather shows a decrease of $Q$ as shown in Figure~\ref{f:SIFigure4}. As pointed out in the main text non saturable attenuation might be caused by heavily asymmetric TLS and modified scattering cross-sections. The influence of individual TLS with deviating energies and tunneling splittings from the assumed contribution might also be increased due to the tiny membrane volume. The unexpected behaviour around $T_\mathrm{mc}=\SI{50}{\milli\kelvin}$ in the temperature dependent data (Figure~\ref{f:SIFigure3}) might be an additional indication for that. 

\begin{figure}
	\centering
	\includegraphics{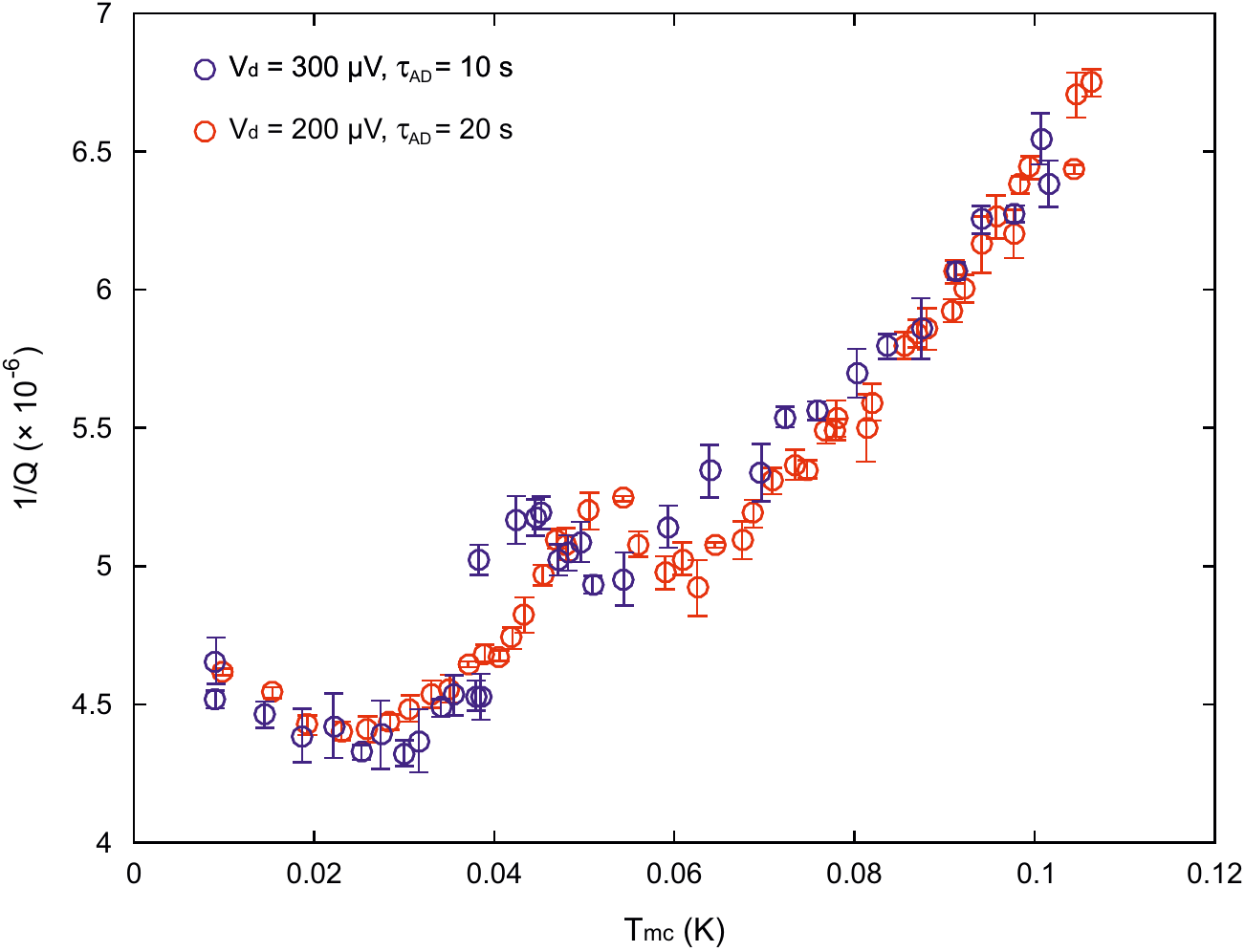}%
	\caption{
		The $1/Q$ measured as function of different drive voltages and measurement times. Both the measurements are consistent within the error bars and were taken coming from low temperature and heating up the sample. Both measurements show data at $\SI{0}{\volt}$ and $P_\mathrm{in}=\SI{0}{\decibel m}$. However, the first measurement (blue data points) uses a $\SI{100}{\micro\volt}$ higher drive of $V_\mathrm{d}=\SI{300}{\micro\volt}$ and $\SI{10}{\micro\second}$ faster time constant of $\tau_\mathrm{AD} =\SI{10}{\micro\second}$. The second measurements (red data points) have been taken at $V_\mathrm{d}=\SI{200}{\micro\volt}$ and an acquisition time constant of $\tau_\mathrm{AD} = \SI{20}{\micro\second}$. 
	}
	\label{f:SIFigure3} 
\end{figure}

\begin{figure}
	\centering
	\includegraphics{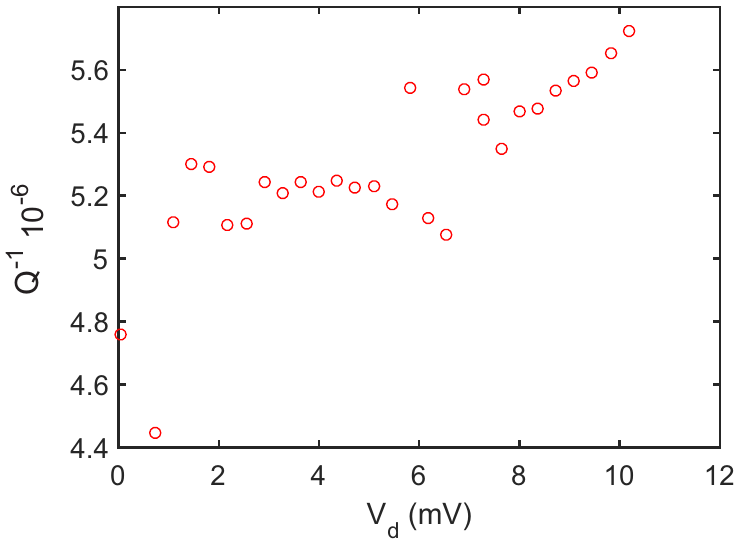}%
	\caption{$1/Q$ measured at base temperature as function of drive voltage. No saturation of the damping with increased drive voltage is observed.
	}
	\label{f:SIFigure4}
\end{figure}
\newpage
\subsection{Effect of thermal expansion on the mechanical resonance frequency}
One possible explanation for the observed discrepancy of the frequency shift between the TLS theory and the measured data above $T_{co}$ in favor of lower frequencies is the influence of the negative thermal expansion of the graphene layers.
Assuming that the mechanical properties of the multilayer are dominated by the graphene based on the higher Youngs modulus one can roughly estimate the frequency shift due to the change in strain of the membrane. 
Using a value of $\alpha = -5\cdot10^{-6}$/K for the thermal expansion coefficient (slightly below room temperature value~\cite{Yoon2011_SI}) and $\Delta\omega/\omega=\sqrt{(1+\alpha\cdot\Delta T)\cdot\omega-\omega}/\omega$ results in the red curve in Fig. \ref{f:SIFigure5}. We point out that neither the expansion of the exact resonator drum was regarded nor a deeper investigation in the actual size of this contribution was conducted and the analysis serves only to give a rough estimation of the effect. 
\begin{figure}
	\centering
	\includegraphics{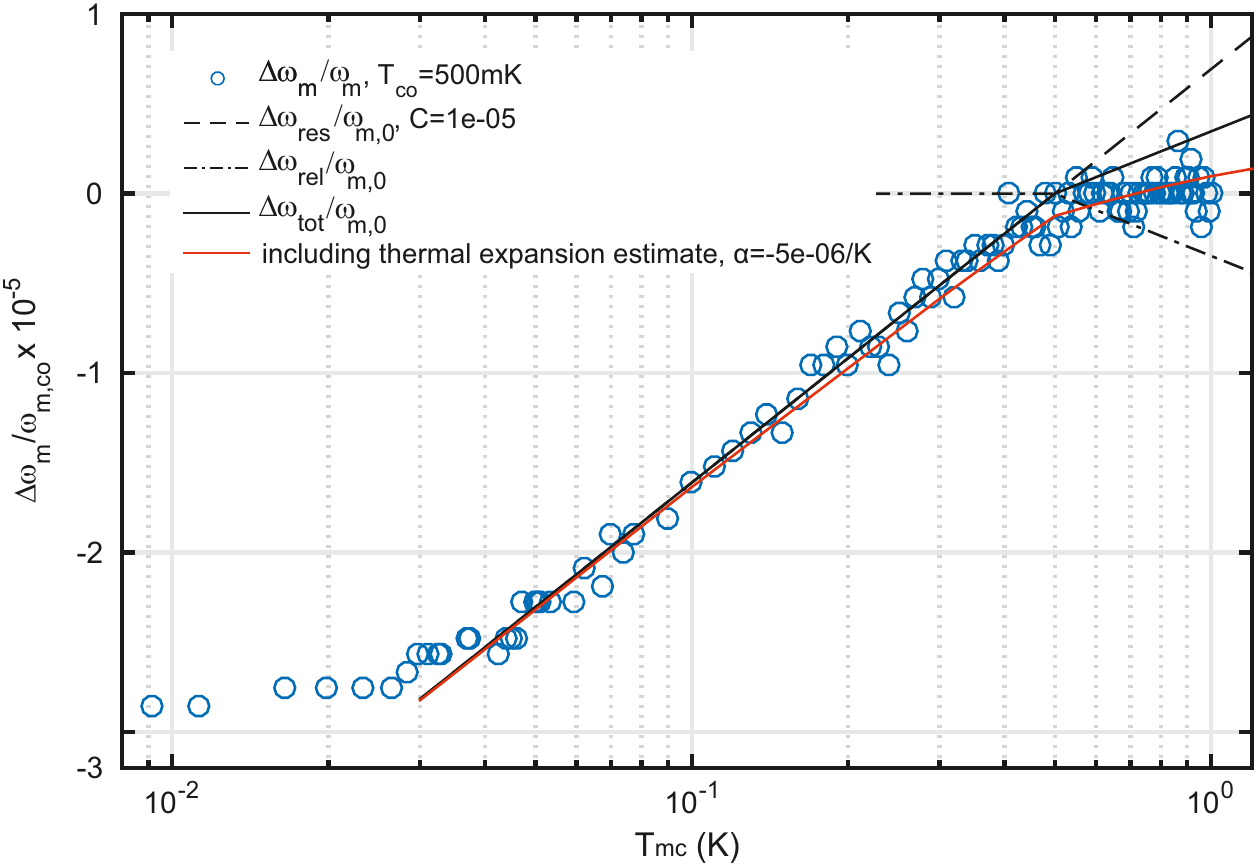}%
	\caption{Temperature dependence of the relative shift in mechanical resonance frequency with temperature. The red curve illustrates the influence of an additional contribution to the frequency shift caused by variation of the strain in the membrane due to the negative thermal expansion of the graphene layers.}
	\label{f:SIFigure5}
\end{figure}
\newpage
\providecommand{\latin}[1]{#1}
\makeatletter
\providecommand{\doi}
{\begingroup\let\do\@makeother\dospecials
	\catcode`\{=1 \catcode`\}=2 \doi@aux}
\providecommand{\doi@aux}[1]{\endgroup\texttt{#1}}
\makeatother
\providecommand*\mcitethebibliography{\thebibliography}
\csname @ifundefined\endcsname{endmcitethebibliography}
{\let\endmcitethebibliography\endthebibliography}{}

}
\end{document}